\documentclass[a4paper, review, 11pt, nonatbib]{elsarticle}
\journal{arXiv}

\usepackage[a4paper,text={16cm,25cm}]{geometry}
\usepackage[labelsep=period]{caption}

\usepackage{array}
\usepackage{booktabs}
\usepackage{amsthm}
\usepackage{amsmath}
\usepackage{amsfonts}
\usepackage{graphicx}
\usepackage{xurl}

\makeatletter
\let\c@author\relax
\makeatother
\usepackage[style=authoryear-comp, backend=biber, uniquename=false, maxnames=99, url=false, giveninits=true, alldates=year, maxcitenames=3, natbib=true, safeinputenc]{biblatex}
\renewbibmacro{in:}{%
  \ifentrytype{article}{}{%
      \printtext{\bibstring{in}\intitlepunct}}}
  \DeclareNameAlias{sortname}{family-given}
\AtEveryBibitem{%
  \clearfield{doi}%
  \clearfield{addendum}%
  \clearfield{eventtitle}%
  \clearfield{issn}%
  \clearfield{url}%
  \clearlist{urldate}%
  \clearlist{language}%
}

\usepackage{longtable,booktabs}
\usepackage{algorithm}
\usepackage{algorithmicx}
\usepackage{algpseudocode}
\usepackage{verbatim}

\usepackage{hyperref}
\hypersetup{
  colorlinks=true,
  linkcolor=blue,
  citecolor=blue,
  urlcolor=blue}

\usepackage{bm}

\usepackage{setspace}
\usepackage[compact]{titlesec}
\titlespacing{\section}{0pt}{0ex}{0ex}
\titlespacing{\subsection}{0pt}{0ex}{0ex}
\titlespacing{\subsubsection}{0pt}{0ex}{0ex}

\usepackage{enumitem}

\begin{document}
\begin{frontmatter}

\title{A Forecaster's Review of Judea Pearl's Causality: Models, Reasoning and Inference, Second Edition, 2009}

\author[address]{Feng Li\corref{acknowledgment}}
\cortext[acknowledgment]{This research was supported by the National Social Science Found of China (22BTJ028).}
\ead{feng.li@cufe.edu.cn}
\address[address]{School of Statistics and Mathematics, Central University of Finance  and Economics, Beijing 102206, China}

\begin{abstract}

  With the big popularity and success of Judea Pearl's original causality book, this review covers the main topics updated in the second edition in 2009 and illustrates an easy-to-follow causal inference strategy in a forecast scenario. It further discusses some potential benefits and challenges for causal inference with time series forecasting when modeling the counterfactuals, estimating the uncertainty and incorporating prior knowledge to estimate causal effects in different forecasting scenarios.

 \end{abstract}


\end{frontmatter}


\section{Introduction and contents}
\label{sec:intro}

With the big popularity and success of Judea Pearl's original causality book in 2000, the second edition published in 2009 further updates and clarifies many details in all ten chapters of the original book. The book has a mixed targeting audience. Readers with basic probability and statistical knowledge would be easy to follow the contents. Readers with nonmathematical backgrounds could skip most of the formulas and still find the concepts and discussion useful. Moreover, each chapter of the second edition elaborates summaries of new developments, discussions and annotated bibliographies for in-depth reading.

The second edition of the book starts with the basics of probabilities and graphs. The causal model is designed with causal networks as oracles. Then the second chapter describes the possibility of learning causal relationships from raw data with a causal discovery framework. The next chapter explores the ways of inferring such relationships by computing the effect of interventions and controlling confounding bias. The inference rule is represented by the \emph{do-calculus} which gives these causal notions a clear empirical interpretation. Chapter 4 extends the effect of a fixed constant $x$ in $do(x)$ with a probabilistic setting of $x$. This is to facilitate the evaluation of the effect of novel actions and policies. Sequences of time-varying actions could also be designed to evaluate the effects with a graphical method. Chapter 5 demonstrates graphical models and the logic of intervention can alleviate the current difficulties in structural equation modeling for causal analysis. Furthermore, Chapter 6 addresses the difficulties encountered when we attempt to define and control confounding by using statistical criteria. And Chapter 7 provides a formal analysis of structure-based counterfactuals. Chapter 8 describes how graphical and counterfactual models can combine to elicit causal information from imperfect experiments. Chapter 9 provides formal semantics for the probability that one event was a necessary or sufficient cause (or both) of another event using counterfactual interpretations. Chapter 10 shows the actual causation that can be formulated in structural model semantics. The last chapter reflects, elaborates, and discusses many philosophical and empirical cases with readers. The epilogue is an introduction to the non-mathematical aspects of causation.

Causality inference has achieved significant advances in the past two decades. With the increase in available data, causal inference with graphic models has been adapted to handle large-scale and high-dimensional datasets. It provides flexible approaches to handling complex data structures and incorporates prior knowledge for causal inference. Various causal discovery and inference algorithms have been developed. These algorithms provide automated methods to infer causal relationships from observational and experimental data. New methods have been developed to address challenges in inferring causality from observational data, such as propensity score matching, and regression discontinuity designs. These techniques aim to overcome confounding biases and establish causal relationships.

The integration of causality and machine learning has received significant attention. Methods that combine causal inference with machine learning techniques, such as causal forest, causal boosting, and causal generative models, have been proposed to leverage the strengths of both fields and enable causal reasoning in predictive models. Interestingly, although many machine learning and deep learning methods have been successfully adopted in the forecast community, there is still relatively little focus on causality for time series forecasting. A shred of evidence is that Pearl nicely collected journal reviews for the causality book in various disciplines \footnote{Available at \url{http://bayes.cs.ucla.edu/BOOK-2K/book_review.html}.} which is a nice addition to the book. It is not surprising that I did not find any related review from a forecasting perspective.

To review the concepts of causality from a forecasting perspective, we consider a typical forecasting scenario where we want to forecast the sales performance of different retail stores over time such as the M5 competition \citep{MakridakisS2022M5Accuracy}. A few questions immediately interest us.

\begin{itemize}
\item Why should forecasters know about causality?
\item How do I know if I have enough data for causal inference?
\item What tools can be directly picked up by forecasters?
\item Are we still suffering from computational challenges?
\end{itemize}

\section{Causality in a forecast scenario}

Business planners may not be satisfied by merely tackling forecast accuracy. A forecaster (or more generally, a prophet) would be expected to provide more insights on top of forecasts. Although not explicitly declared in Pearl's causality book, it is assumed the data are always available. Our aforementioned scenario required the dataset with information on various factors such as store size, location, promotional activities, and historical sales data. Unfortunately, not every time series dataset contained such detailed information. It is not uncommon that only historical sales data are available at hand. Historically, forecasters were in difficulty involving causality analysis. Nowadays, the situation is much better because the external variables are easily collected. For example, the M5 competition \citep{MakridakisS2022M5Accuracy} dataset includes many external variables as features. Machine learning algorithms could work on such features to improve forecasting performance.

Now we could construct a causal model that represents the relationships among these variables. We can use directed acyclic graphs (DAGs) to visually represent the causal structure. For example, we may have arrows from store size, location, and promotional activities to sales performance, indicating their potential causal influence.

With the causal model and data in place, we can use the \emph{do-calculus} to specify interventions. When we examine the causal effect of store size on sales performance, we can use the \emph{do-operator} and write it as ``\emph{do}(store size)''. This signifies that we are considering a hypothetical scenario where we actively intervene and change the store size, regardless of its original causal mechanisms. By applying the \emph{do-calculus}, we can evaluate the causal effect of the intervened variable (in this case, store size) on the outcome variable (sales performance). We can compare the outcomes under the intervention ``\emph{do}(store size)'' to the outcomes without the intervention to estimate the causal effect. This allows us to isolate the impact of store size on sales performance while holding other factors constant.

The causal analysis also identifies valid adjustment sets for estimating causal effects with the back-door criterion. It helps determine which variables need to be controlled for in order to obtain an unbiased estimate of the causal effect of an intervention. We could utilize the back-door path from the causal variable (e.g., store size) to the outcome variable (e.g., sales performance) that contains an arrow entering the causal variable. In the retail store example, a potential back-door path could be through the variable ``promotional activities.'' The criterion requires that all back-door paths between the causal variable and the outcome are blocked.

An adjustment set is a set of variables that, when conditioned on (adjusted for), blocks all back-door paths. To satisfy the back-door criterion, the adjustment set must contain variables that are ancestors of the causal variable (store size) in the causal graph but not descendants of the causal variable or intermediates on the causal path between the causal variable and the outcome. The adjustment set should not include variables that are affected by the intervention (store size). Including post-treatment variables in the adjustment set can introduce bias into the estimation of the causal effect.

In practice, identifying a valid adjustment set may require domain knowledge, subject expertise, and careful consideration of the causal relationships among variables. Techniques such as propensity score matching, stratification, or regression adjustment can be used to implement the adjustment set and estimate the causal effect while satisfying the back-door criterion.

 If we further assess the impact of promotional activity on sales performance. We can construct a counterfactual scenario where the promotional activity was not conducted (denoted as ``\emph{do}(promotional activity = false)''). This allows us to compare the actual sales performance with the hypothetical sales performance that would have occurred if the promotional activity had not taken place.  If the actual sales performance is higher than the counterfactual outcome, it suggests a positive causal effect, indicating that the promotional activity had a beneficial impact on sales. Conversely, if the actual sales performance is similar or lower than the counterfactual outcome, it indicates a minimal or negative causal effect.

\section{Discussion}

In general, causality analysis requires more data than traditional forecasting methods because it aims to identify causal relationships between variables. While traditional forecasting methods may only require a few variables to make predictions, causal models typically require more variables and a deeper understanding of the underlying mechanisms driving the data. This can make it difficult for forecasters who have limited data or data that is noisy or incomplete or not accessible due to privacy concerns. A possible direction is to design a time series data collecting protocol so that time series data can be properly used for causal analysis.

The book demonstrates the advances of Pearl's causality framework with Bayesian graphs, but in general causal inference also involves more complex computations than traditional forecasting methods. Graphical models and DAGS, which are commonly used in Pearl's framework to represent causal relationships, can be computationally intensive to build and analyze. Causal inference algorithms, such as the \emph{do-calculus} and backdoor criterion, can also be complex to apply and may require specialized software. This could still be a challenge for forecasters without a strong background in statistics and computer science.

Any causality framework relies on several assumptions, such as the assumption of faithfulness and the requirement for no unmeasured confounding variables in Pearl's causality. Violations of these assumptions can lead to inaccurate forecasts and may be difficult for forecasters to identify and address. Additionally, the framework may not be directly applicable to all forecasting scenarios, particularly when more machine learning models are getting attracted to the forecasting domain. See \citet{ScholkopfB2022CausalityMachine} for possible solutions.

Computer software is also essential for causal analysis. I found an open-source Python library ``Causalnex'' \citep{BeaumontP2021CausalNexToolkit} a good companion. It aims for building, testing, and applying Bayesian networks for causal modeling and inference. The library includes tools for time series modeling, with support for incorporating external variables, handling missing data, and performing counterfactual analysis. Causalnex also provides visualization tools for exploring causal relationships and model outputs.

Overall, Pearl's causality framework offers the ladder for time series forecasting and Bayesian graphs that allows for the explicit modeling of counterfactuals, uncertainty estimation, and incorporating prior knowledge to estimate causal effects in different scenarios. There exist issues including the definition of identifiability and the choice of priors in both low and high dimensional regimes \citep{LiF2023BayesianCausal}. A plausible approach is to consider the causal machine learning methods \citep{KaddourJ2022CausalMachine, LiuY2022RobustAdaptive}. Furthermore,  there are many avenues for causal inference in forecasting with generative models, given the wide applicability of simulations in forecasting models. Forecasters must carefully consider the data requirements, computational complexity, need for domain expertise, interpretability, and assumptions of the framework before adopting it. Addressing these challenges may require specialized expertise, computational resources, and careful consideration of the specific forecasting problem at hand.

\printbibliography
\end{document}